\documentclass[conference]{IEEEtran}
\usepackage[latin9]{inputenc}
\usepackage{amsmath, amssymb}
\usepackage{graphicx}

\ifCLASSINFOpdf
\else
\fi
\usepackage{amsmath}
\usepackage{stfloats}
\begin{document}
%
\title{Group Membership Verification with Privacy:\\ Sparse or Dense?}

\author{\IEEEauthorblockN{Marzieh Gheisari}
\IEEEauthorblockA{Univ Rennes, Inria, CNRS, IRISA \\France}
\and
\IEEEauthorblockN{Teddy Furon}
\IEEEauthorblockA{Univ Rennes, Inria, CNRS, IRISA\\
France
}
\and
\IEEEauthorblockN{Laurent Amsaleg}
\IEEEauthorblockA{Univ Rennes, Inria, CNRS, IRISA\\France}
}


%



\newcommand{\seq}[1]{\mathbf{#1}}
\newcommand{\fun}[1]{\mathsf{#1}}
\newcommand{\set}[1]{\mathcal{#1}}

\def \Y {\seq{Y}}
\def \X {\seq{X}}
\def \Q {\seq{Q}}
\def \N {\seq{N}}
\def \Hyp {\mathcal{H}}
\def \iid {\mathsf{i.i.d.}}
\def \a {\fun{a}}
\def \s {\fun{s}}
\def \hash {\fun{h}}
\def \V {\vec{V}}
\def \W {\vec{W}}
\def \U {\vec{U}}
\def \Z {\vec{Z}}
\def \v {\vec{v}}
\def \u {\vec{u}}
\def \w {\vec{w}}
\def \z {\vec{z}}
\def \b {\mathbbm{b}}
\def\un{{\mathbbm{1}}}
\def \Rx {\mathcal{X}}
\def \Rq {\mathcal{Q}}
\def \Vs {\mathcal{V}}
\def \real {\mathbb{R}}
\def \thetab {\boldsymbol{\theta}}
\def \Pr {\mathbb{P}}
\def \Pfp {P_{\mathsf{fp}}}
\def \Pfn {P_{\mathsf{fn}}}
\def \Efp {E_{\mathsf{fp}}}
\def \Alp {\mathcal{X}}
\def \Alps{|\mathcal{X}|}
\def \Alpy {\mathcal{Y}}
\def \Alpt {\mathcal{T}}
\def \ie {\textit{i.e.\,}}
\def \fC {\fun{C}}
\def \fV {\fun{V}}
\def \fS {\fun{S}}
\def \thetab {\boldsymbol{\theta}}
\def \vx {\vec{x}}
\def \vq {\vec{q}}

\maketitle
\def \Pfn {P_{\mathsf{fn}}}
\def \Pfp {P_{\mathsf{fp}}}
\def\ie{{\it i.e.}}

\begin{figure}[b]
\parbox{\hsize}{\em
WIFS`2019, December, 9-12, 2019, Delft, Netherlands.
978-1-7281-3217-4/19/\$31.00 \ \copyright 2019 European Union.
}\end{figure}

\begin{abstract}
Group membership verification checks if a biometric trait corresponds to one member of a group without revealing the identity of that member. 
 Recent contributions provide privacy for group membership protocols through the joint use of  two mechanisms: quantizing templates into discrete embeddings, and aggregating several templates into one group representation.

However, this scheme has one drawback: the data structure representing the group has a limited size and cannot recognize noisy  query when many templates are aggregated. Moreover, the sparsity of the embeddings seemingly plays a crucial role on the performance verification.

This paper proposes a mathematical model for group membership verification allowing to reveal the impact of sparsity on both security, compactness, and verification performances.
This models bridges the gap towards a Bloom filter robust to noisy queries.  
It shows that a dense solution is more competitive unless the queries are almost noiseless. 
\end{abstract}


%
\IEEEpeerreviewmaketitle

\section{Introduction}
Group membership verification is a procedure checking whether an item or an individual is a member of a group. If membership is positively established, then an access to some ressources (a building, a file, \ldots) is granted; otherwise the access is refused. This paper focuses on \emph{privacy preserving} group membership verification procedures where members must be distinguished from non-members, but where the members of a group should not be distinguished one another.

To this aim, a few recent contributions have proposed to rely on the aggregation and the embedding of several distinctive templates into a unique and compact high dimensional feature representing the members of a group~\cite{Gheisari2019icassp, Gheisari_2019_CVPR_Workshops}. It has been demonstrated that this allows a good assessment of the membership property at test time. It has also been shown that this provides privacy and security.  Privacy is enforced because it is impossible to infer from the aggregated feature which original distinctive template matches the one used to probe the system. Security is preserved since nothing meaningful leaks from embedded data~\cite{Razeghi2017wifs,Razeghi2018icassp}.

\cite{Gheisari2019icassp} and~\cite{Gheisari_2019_CVPR_Workshops}, however face severe limitations. Basically, it seems impossible to create features representing groups having many members. In this case, the probability to identify true positives vanishes and the false negative rate grows accordingly. Furthermore, the robustness of the matching procedure fades and becomes unable to absorb even the smallest amount of noise that inherently differentiate the enrolled template of one member and the template captured at query time for this same member. In contrast, features representing only few group members are robust to noise and cause almost no false negatives. A detailled analysis of~\cite{Gheisari2019icassp} and~\cite{Gheisari_2019_CVPR_Workshops} suggests that these limitations originate from the sparsity level of the features representing group members.

This paper investigates the impact of the sparsity level of the high dimensional features representing group members on the quality of (true positive) matches and on their robustness to noise. It shows it is possible to trade compactness and sparsity for better security or better verification performance.

Sect.~\ref{sec:Setup}  first considers the aggregation of discrete random sequences, and models this compromise with information theoretical tools. Sect.~\ref{sec:Binary} applies this viewpoint to binary random sequences and shows that the noise on the query has an impact depending on the sparsity of the sequences. Sect.~\ref{sec:Real} bridges the gap between the templates, \ie\ real $d$-dimensional vectors, and the discrete sequences considered in the previous sections. Sect.~\ref{sec:Experiments} gathers the experimental results for a group membership verification based on faces.


\section{Discrete Sequences}
\label{sec:Setup}
This section considers the problem of creating a representation $\Y$ of a group of $n$ sequences $\{\X_{1}, \ldots,\X_{n}\}$, whose use is to test whether a query sequence $\Q$ is a noisy version of one of these $n$ original sequences. This test is done at query time when the original sequences are no longer available and all that remains is the representation  $\Y$.

The sequences are elements of $\Alp^{m}$ where $\Alp$ is a finite alphabet of cardinality $|\Alp|$, say $\Alp:=\{0,1,\ldots,\Alps-1\}$.
The sequence follows a statistical model giving a central role to the symbol $0$.
The symbols of the sequences are independent and identically distributed with
\begin{equation}
\label{eq:SparseDef}
\Pr(X = s) = \begin{cases}
1 - p(\Alps-1) & \text{if } s =0\\
p & \text{otherwise}
\end{cases}
\end{equation}
for $p\in(0,1/\Alps]$.
Sparsity means that probability $p$ is small, density means that $p$ is close to $1/\Alps$ so that $X$ is uniformly distributed over $\Alp$.

\subsection{Structure of the group representation}
We impose the following conditions on the aggregation $\a(\cdot)$ computing the group representation $\Y = \a(\X_{1},\ldots,\X_{n})$:
\begin{itemize}
\item $\Y$ is a discrete sequence of the same length $\Y\in\Alpy^{m}$,
\item Symbol $Y(i)$ only depends on symbols $\{X_{1}(i),\ldots,X_{n}(i)\}$,
\item The same aggregation is made index-wise: with abuse of notation, $Y(i) = \a(X_{1}(i),\ldots,X_{n}(i))$, $\forall i\in[m]$,
\item $Y(i)$ does not depend on any ordering of the set $\{X_{1}(i),\ldots,X_{n}(i)\}$,
\end{itemize}
These requirements are well known in traitor tracing and group testing as they usually model the collusion attack or the test results over groups. Here, they simplify the analysis reducing the problem to a single letter formulation where index $i$ is dropped involving symbols $\{X_1,\ldots,X_n\}$, $Y$ and $Q$.

These conditions motivate a 2-stage construction.
The first stage computes the type (a.k.a. histogram or tally) $T$ of the symbols $\{X_{1},\ldots,X_{n}\}$.
Denote by $\Alpt_{|\Alp|,n}$ the set of possible type values.
Its cardinality equals $|\Alpt_{|\Alp|,n}| = \binom{n+\Alps-1}{\Alps-1}$ which might be too big.
The second stage applies a surjective function $\fun{r}:\Alpt_{|\Alp|,n}\to\Alpy$, where $\Alpy$ is a much smaller set.


\subsection{Noisy query}
\label{sec:Noisy}
At enrollment time, the system receives $n$ sequences, aggregates them into the compact representation $\Y$, and then forgets the $n$ sequences.
At query time, the system receives a new sequence $\Q$ conforming with one of the following hypotheses:
\begin{itemize}
\item $\Hyp_{1}$: $\Q$ is a noisy version of one of the enrolled sequences.
Without loss of generality, $\Q = \X_{1}+ \N$. 
\item $\Hyp_{0}$: $\Q = \X_0 + \N$, where $\X_0$ shares the same statistical model but it is independent of $\{\X_{1},\ldots,\X_{n}\}$.
\end{itemize}
We model the source of noise (due to different acquisition conditions) by a discrete communication channel. 
It is defined by function $\fun{W}:\Alp\times\Alp\to[0,1]$ with $\fun{W}(q|x):= \Pr(Q = q|X=x)$.
We impose some symmetry w.r.t. the symbol $0$: $\fun{W}(s|0) = \eta_{0}$ and
$\fun{W}(0|s)=\eta_{1}$, $\forall s\in\Alp \backslash\{0\}$.

At query time, the system computes a score $S = \s(\Q,\Y)$ and compares to a threshold:
hypothesis $\Hyp_{1}$ is deemed true if $S\geq\tau$.
This test leads to two probabilities of error:
\begin{itemize}
\item $\Pfp(n,m)$ is the probability of false positive: 
$\Pfp(n,m) := \Pr(S\geq\tau|\Hyp_{0})$. 
\item $\Pfn(n,m)$ is the probability of false negative:
$\Pfn(n,m) := \Pr(S<\tau|\Hyp_{1})$.
\end{itemize}
The emphasis on $(n,m)$ is natural. It is expected that:
i) the more sequences are aggregated, the less reliable the test is,
ii) the longer the sequences are, the more reliable the test is.

\subsection{Figures of merit $(\fC, \fS, \fV)$}
The section presents three information theoretic quantities (expressed in nats) measuring the performances of the scheme.
The first two depends on the statistical model of $X$ (especially~$p$) and the aggregation mechanism $\a$.
The last one depends moreover on the channel. 

\subsubsection{Compactness $\fC$}
The compactness of the group representation is measured by the entropy $\fC:= H(Y)$. 
It roughly means that the number of typical sequences $\Y$ scales exponentially as $e^{mH(Y)}$, which can be theoretically compressed to the rate of $H(Y)$ nats per symbol.

\subsubsection{Security $\fS$}
We consider an insider aiming at disclosing one of the $n$ enrolled sequences.
Observing the group representation $\Y$, its uncertainty is measured by the equivocation $\fS:=H(X|Y)$.  
This means that the insider does not know which of the $e^{mH(X|Y)}$ typical sequences the enrolled sequences are. 

\subsubsection{Verification $\fV$}
In our application, the requirement of utmost importance is to have a very small probability of false positive.  
We are interested in an asymptotical setup where $m\to+\infty$. This motivates the use of the false positive error exponent as a figure of merit:
\begin{equation}
\Efp(n) := \lim_{m\to+\infty} -\frac{1}{m}\log \Pfp(n,m).
\end{equation} 
If $\Efp(n)>0$, it means that $\Pfp(n,m)$ exponentially vanishes as $m$ becomes larger.
The theory of test hypothesis shows that $\Efp(n)$ is upper bounded by the mutual information $\fV := I(Y;Q)$ where $Q$ is a symbol of the query sequence, \ie\ a noisy version of $X_1$.
It means that the necessary length for achieving the requirement $\Pfp(n,m)<\epsilon$ is~\cite{Shannon:1959qv}
\begin{equation}
\label{eq:Shannon}
m \geq \frac{-\log\epsilon}{\fV}.
\end{equation}


\subsection{Noiseless setup}
The bigger $\fV$ and $\fS$, the better the performance in terms of verifiability and security.
Yet, they can not be both big at the same time. 
The noiseless case when the channel introduces no error and $Q=X$ simply illustrates the trade-off:
\begin{eqnarray}
\fV &\leq& \fC\\
\fV + \fS &=& H(X), \label{eq:VSH}
\end{eqnarray}
with $H(X)=-\log p_0 + (1-p_0)\log\frac{p}{p_0}$ and $p_0:=\Pr(X=0)$~\eqref{eq:SparseDef}.
For a given $\Alps$, $H(X)$ is maximised by the dense solution:
$H(X) \leq \log\Alps$ with equality for $p=1/\Alps$. 


\section{Binary alphabet}
\label{sec:Binary}
 This section explores the binary case where $\Alp = \{0,1\}$.
 We first set the surjection as the identity function s.t. $Y = T$.
 Then, the impact of the surjection is investigated.
 
 \subsection{Working with types}
 In the binary case, there are $n+1$ type values.
 There can be uniquely labelled by the number of symbols `1' in $\{X_1,\ldots,X_n\}$, \ie\ $T = \sum_{i=1}^n X_i\sim\mathcal{B}(n,p)$. 

\subsubsection{Verification}
In the noiseless case, after some rewriting:
\begin{equation}
\fV = \fun{h}(p) - \sum_{t = 0}^n \Pr(T = t) \fun{h}\left(\frac{t}{n}\right), 
\end{equation}
with
$\fun{h}(p):=-p\log(p)-(1-p)\log(1-p)$, the entropy of a Bernoulli r.v. $\mathcal{B}(p)$.
If $p=1/2$ and $n$ is large:
\begin{equation}
\fV = \frac{1}{2n} + o\left(\frac{1}{n}\right).
\label{eq:Vdense}
\end{equation}
This is not the maximum of this quantity. For large $n$, the best option is to set
\begin{equation}
p=\frac{\alpha}{n},\quad\fV = \frac{\beta}{n} + o\left(\frac{1}{n}\right),
\end{equation}
with $\alpha = 1.338$ and $\beta = 0.580$.
This was proven in the totally different application of traitor tracing~\cite[Prop.~3.8]{Laarhoven:2015aa}.

This section outlines two setups: the dense setup where $p=1/2$, and the sparse setup where $p$ goes to $0$ when more sequences are packed in the group representation. Both setups share the asymptotical property that $\fV\approx \kappa/n$ for large $n$.
According to~\eqref{eq:Shannon}, we can pack a big number $n$ of sequences into one group representation provided that  their length $m$ scales proportionally to $n$. 

\subsubsection{Compactness}
The figure of merit for compactness for types is just $\fC = H(T)$ where $T$ follows a binomial distribution: $T\sim\mathcal{B}(n,p)$. In the dense setup $p=1/2$, the binomial distribution is approximated by a Gaussian distribution $\mathcal{N}(n/2;n/4)$ providing:
\begin{equation}
\label{eq:Cgauss}
\fC = \frac{1}{2}\log\left(\frac{\pi e n}{2}\right) + O \left(\frac{1}{n}\right).
\end{equation}
In the sparse setup $p = \alpha/n$, the binomial distribution is approximated by a Poisson distribution $\mathcal{P}(\alpha)$~\cite{Boersma:1988aa}:
\begin{equation}
\fC \approx \alpha(1 - \log(\alpha)) + e^{-\alpha}\sum_{j=0}^{+\infty} \frac{\alpha^j\log(j!)}{j!}.
\end{equation}
This shows that the types are not compact in the dense setup; It approximatively remains constant in the sparse setup.


\subsubsection{Security}
Thanks to~\eqref{eq:VSH}, we only need to calculate $H(X) = \fun{h}(p)$.
In the dense setup, $H(X) = \log(2)$ and $\fS$ converges to $H(X)$ as $n$ increases.
Merging into a single representation protects an individual sequence. 
If sparse,
\begin{equation}
H(X) = \frac{\alpha}{n}\left(1 - \log\frac{\alpha}{n}\right) + o\left(\frac{1}{n}\right).
\end{equation}
Therefore, $\fS$ converges to zero as $n$ increases, contrary to the dense setup.
It might be more insightful to see that the ratio of uncertainties before and after observing $T$, \ie\ $H(X)/H(X|T)$, converges to 1 in both cases. Merging does provide some security but sparsity is more detrimental.

\subsection{Adding a surjection}
\label{sub:surjection}
\begin{figure}[tbp]
\begin{center}
\includegraphics[width = 0.9\columnwidth]{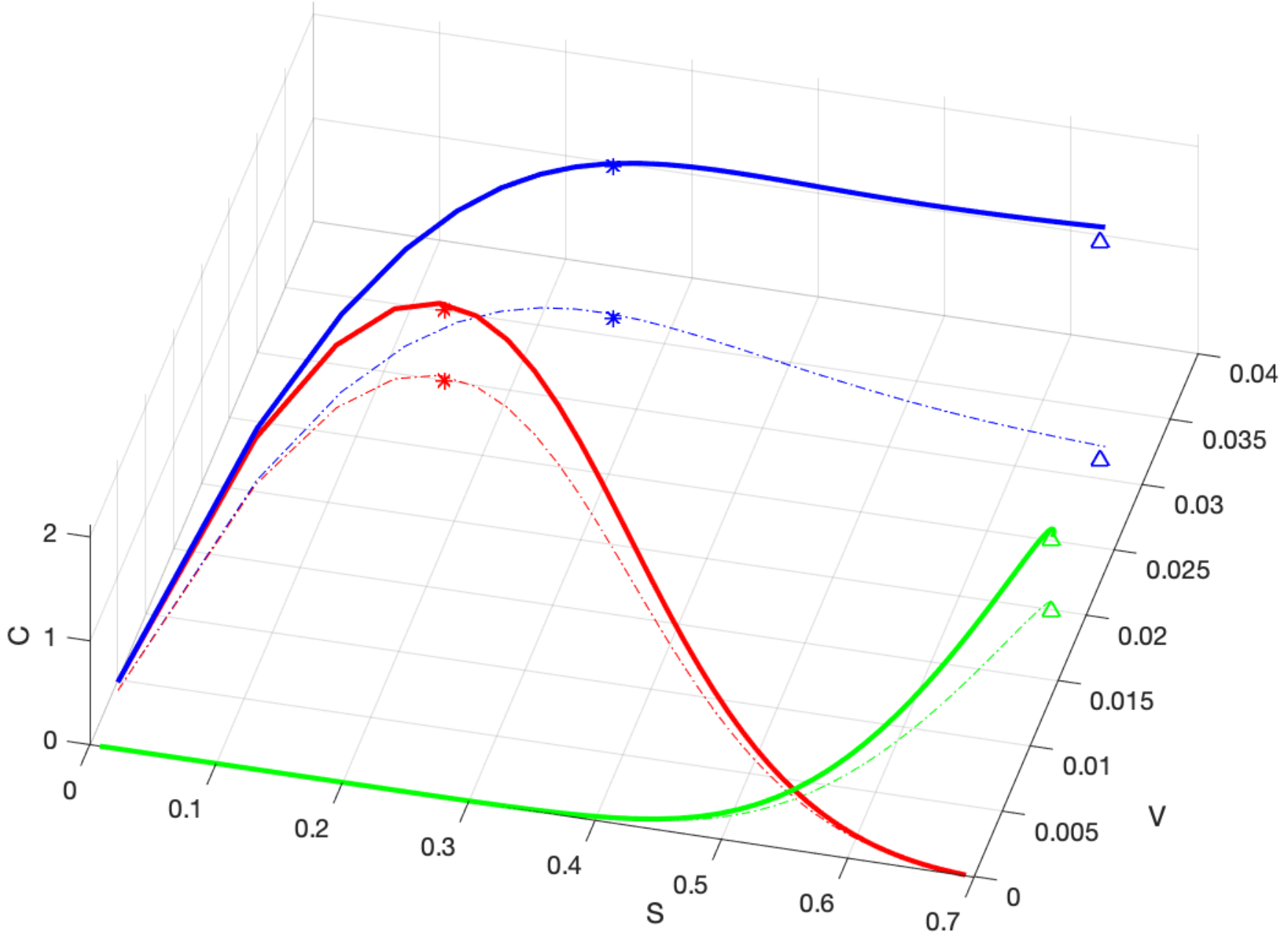}
\caption{The trade-off $(\fS,\fV,\fC)$ for $\Alp=\{0,1\}$, $n=16$, $Y = T$ (blue), $Y=\fun{r}(T)$ for `All-1' (red) and majority vote (green). Dashed plot represents the projection onto $\fC=0$. Triangles and stars summarize results~\eqref{eq:Vdense} to~\eqref{eq:Vsparse}.}
\label{fig:tradeoff}
\vspace{-12pt}
\end{center}
\end{figure}

The motivation of the surjection onto a smaller set $\Alpy$ is to bound $\fC$ as $\fC\leq \log|\Alpy|$, $\forall n$.
The Markov chain $Q\rightarrow X_1\rightarrow T\rightarrow Y$ imposes that $\fV\leq I(T;Q)$.
The surjection thus provoques a loss in verification as depicted in Fig.~\ref{fig:tradeoff}.

App.~\ref{app2} shows that for $|\Alpy| = 2$, this loss is minimized for:
\begin{equation}
\fun{r}(t) =
\begin{cases}
0 & \text{if } t < t_p \\
1 & \text{otherwise}
\end{cases}
\end{equation}
where $t_p$ is a threshold depending on $p$.
In the dense setup, $t_p = n/2$ and the surjection corresponds to a majority vote collusion in traitor tracing (a threshold model in group testing).
Hence, by~\cite[Prop.~3.4]{Laarhoven:2015aa}:
\begin{equation}
\fV = \frac{1}{n\pi} + o\left(\frac{1}{n}\right).
\end{equation}
In the sparse setup $t_p = 1$ which corresponds to an `All-1' attack in traitor tracing (a the perfect model in group testing).
Then the best option is to set $p = \log(2)/n$ and~\cite[Prop.~3.3]{Laarhoven:2015aa}:
\begin{equation}
\fV = \frac{(\log(2))^2}{n} + o\left(\frac{1}{n}\right).
\label{eq:Vsparse}
\end{equation}
From~\eqref{eq:Shannon}, the necessary length is $m \geq -n\log(\epsilon)/(\log(2))^2$.

The main property $\fV\approx \kappa / n$ still holds but
the surjection lowers $\kappa$ from $0.5$ to $0.32$ (dense), from $0.84$ to $0.48$ (sparse).
The sparse setup is still the best option w.r.t. $\fV$.

\subsection{Relationship with the Bloom filter}
A Bloom filter is a well-known data structure $\Y\in\{0,1\}^m$ designed for set membership, embedding items to be enrolled into $\Y$ thanks to $k$ hash functions.
Its probability of false negative is exactly $0$, whereas the probability of false positive is not null.
The number of hash functions minimizing $\Pfp(n,m)$ is $k = \lfloor \log(2)m/n\rfloor$.
Then, the necessary length to meet a required false positive level $\epsilon$ is
$m \geq -n\log(\epsilon)/(\log(2))^2$.

These numbers show the connection with our scheme~\eqref{eq:Vsparse}. At the enrollment phase, the hash functions indeed associate to the $j$-th item a binary sequence $\X_j$ indicating which bits of $\Y$ have to be set.
This sequence is indeed sparse with $k/m\approx \log(2)/n$.
The necessary length is the same.
Indeed, the enrollment phase of a Bloom filter is nothing more than the `All-1' surjection.


The only difference resides in the statistical model. There is at most $k$ symbols `1' in sequence $\X_j$ whereas, in our model, that follows a binomial distribution $\mathcal{B}(m,p)$. Yet, asymptotically as $m\to\infty$, by some concentration phenomenon, the two models get similar. This explains why we end up with similar optimal parameters.
Yet, the Bloom filter only works when the query object is exactly one enrolled item, whereas the next section shows that our scheme is robust to noise. 

\section{Real vectors}
\label{sec:Real}
This section deals with real vectors: $n$ vectors to be enrolled $\{\vx_1,\ldots,\vx_n\}\subset\mathbb{R}^d$, and the query vector $\vq\in\mathbb{R}^d$. All have unit norm. An embedding mechanism $\fun{E}:\mathbb{R}^d\to\Alp^m$ makes the connection with the previous section. As in~\cite{Andoni:2015aa}, this study models the embedding as a probabilistic function.
\subsection{Binary embedding}
For instance, for $\Alp = \{0,1\}$, a popular embedding is:
\begin{equation}
\label{eq:Embedding}
X(i) = [\vx^\top \vec{U}_i > \lambda_x], \forall i\in[m]
\end{equation}
where $\vec{U}_i\stackrel{i.i.d.}{\sim}\mathcal{N}(\vec{0}_d,I_d)$.
This in turn gives i.i.d. Bernoulli symbols $\{X(i)\}$ with $p = 1 - \Phi(\lambda_x)$ if $\|\vx\|=1$.

At the query time, the embedding mechanism uses the same random vectors but a different threshold:
\begin{equation}
Q(i) = [\vq^\top \vec{U}_i > \lambda_q], \forall i\in[m].
\end{equation}

Under $\Hyp_1$, suppose that $\vq^\top\vx_1 = c<1$. This correlation defines the channel $X\to Q$ with the error rates:
\begin{eqnarray}
\eta_0 &=& \Pr(\vq^\top \vec{U}>\lambda_q|\vx^\top \vec{U}\leq\lambda_x),\\
\eta_1 &=& \Pr(\vq^\top \vec{U}\leq\lambda_q|\vx^\top \vec{U}>\lambda_x).
\end{eqnarray}
The error rate $\eta_0$ has the expression (and similarly for $\eta_1$):
\begin{equation}
\label{eq:eta0}
\eta_0 = 1 - \frac{1}{(1-p)\sqrt{2\pi}}\int_{-\infty}^{\lambda_x} \Phi\left(\frac{\lambda_q - cx}{\sqrt{1-c^2}}\right) e^{-\frac{x^2}{2}} dx.
\end{equation}
\subsection{Induced channel}
\begin{figure}[htbp]
\begin{center}
\includegraphics[width=0.9\columnwidth]{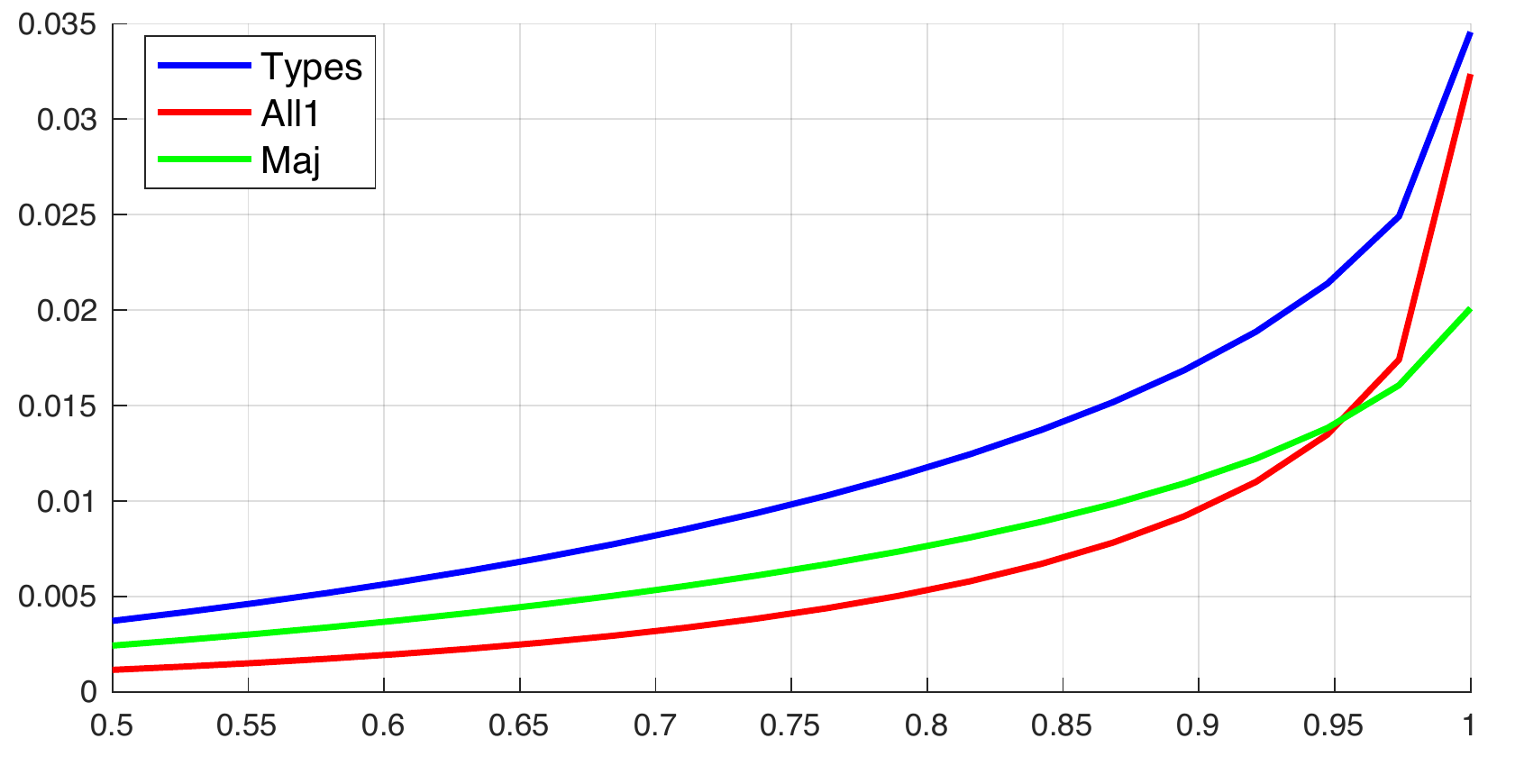}
\caption{$\fV$ as a function of correlation $c$, $d=256$, $n = 15$.}
\label{fig:MIcor}
\end{center}
\end{figure}

For this embedding, the parameters $(\lambda_x, \lambda_q, c,d)$ for the vectors define the setup $(p,\eta_0,\eta_1)$ for the sequences. It is a priori difficult to find the best tuning $(\lambda_x,\lambda_q)$.
For a fixed $\lambda_x$, $\eta_0$ decreases with $\lambda_q$ while $\eta_1$ increases.
App.~\ref{app3} reveals that $\fV$ is sensitive to $\eta_0$ especially with the `All-1' surjection of the sparse solution.
Fig.~\ref{fig:MIcor} shows indeed that the dense solution $(\lambda_x,\lambda_q)=(0,0)$ is more robust, unless $c$ is very close to 1. Here, we enforce a surjection (identity, All-1, or majority vote) and make a grid search to find the optimum $(\lambda_x,\lambda_q)$ for a given $c$. It happens that these parameters are better set to 0, \ie\ dense solution, for the identity and majority vote. As for the `All-1' surjection, we observe that $\lambda_x$ is s.t. $p\approx 1/n$ and $\lambda_q$ is slightly bigger than $\lambda_x$ to lower $\eta_0$. Yet, this sparse solution is not as good as the dense solution unless $c$ is close to 1, \ie\ the query vector is very close to the enrolled vector.

This observation holds only for the embedding function~\eqref{eq:Embedding}.
Hashing functions less prone to error $\eta_0$ may exist.  
  
\section{Experimental work}
\label{sec:Experiments}
\begin{figure*}[tb]%
	\vspace{-12pt}	
	\centering
	\includegraphics[height = 6cm,width=0.95\linewidth]{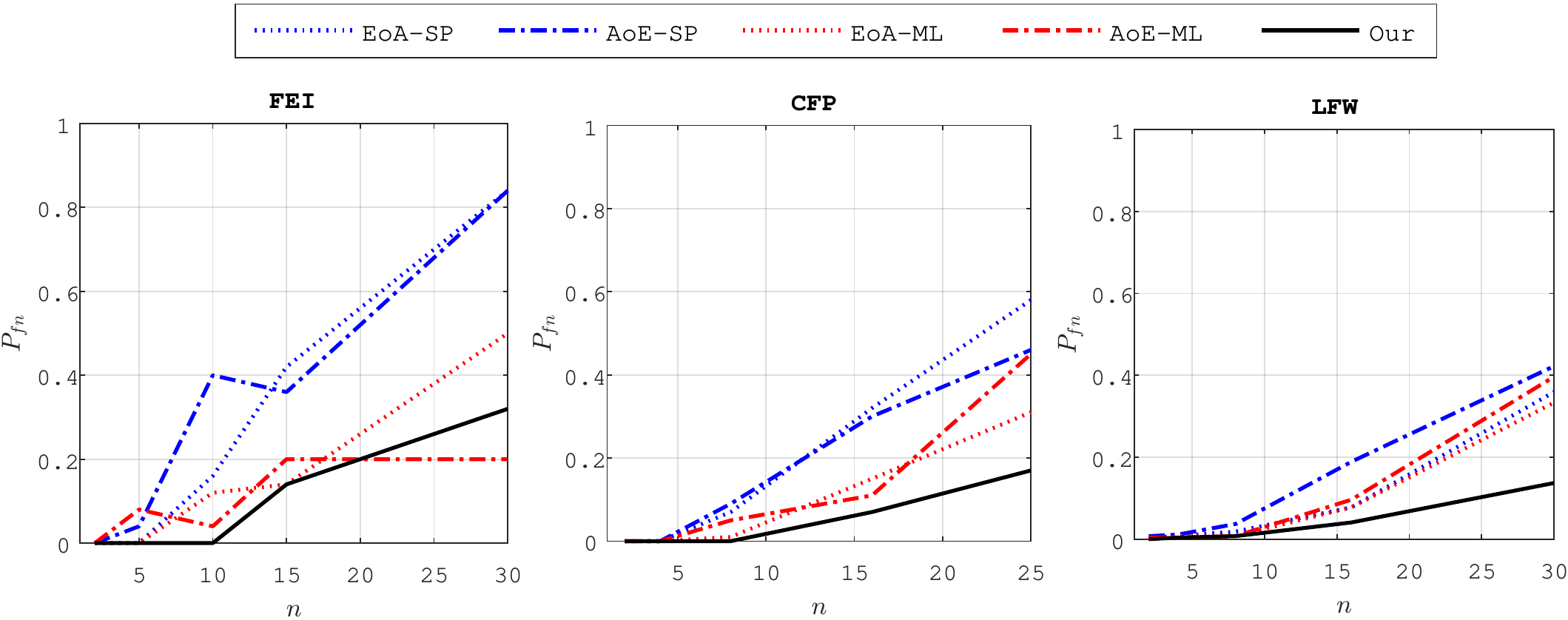}
	\caption{Verification performance $\Pfn@\Pfp=0.05$ \textit{vs.} group size $n$ for the baselines (see Sect.~\ref{fig:exp1}) and our scheme.}
	\label{fig:baseline}%
	\vspace{-12pt}
\end{figure*}

We evaluate our scheme with face recognition.  Face images are coming from LFW~\cite{huang2008labeled}, CFP~\cite{sengupta2016frontal} and FEI~\cite{thomaz2010new} databases.
For each dataset, $N$ individuals are enrolled into random groups.
There is the same number $N_q$ of positive and negative (impostors) queries.
\paragraph*{Labeled Faces in the Wild}
These are pictures of celebrities in all sort of viewpoint and under an uncontrolled environment.
We use pre-aligned LFW images.
The enrollment set consists of $N = 1680$ individuals with at least two images in the LFW database.
One random template of each individual is enrolled in the system, playing the role of $\vx_{i}$.
Some other $N_q = 263$ individuals were randomly picked in the database to play the role of impostors. 

\paragraph*{Celebrities in Frontal-Profile}
These are frontal and profile views of celebrities taken in an uncontrolled environnement.
We only use $N = 400$ frontal images enrolled in the system.
The impostor set is a random selection of $N_q = 100$ other individuals.
\paragraph*{Faculdade de Engenharia Industrial}
The FEI database contains images in frontal view in a controlled environnement. 
We use pre-aligned images.
There are $200$ subjects with two frontal images (one with a neutral expression and the other with a smiling facial expression).
The database is created by randomly sampling $N = 150$ individuals to be enrolled, and $N_q=50$ impostors.
\subsection{Experimental Setup}

Face descriptors are obtained from a \textit{pre-trained} network based on VGG-Face architecture followed by PCA~\cite{parkhi2015deep} .
FEI corresponds to the scenario of employees entering in a building with face recognition, whereas CFP is more difficult, and LFW even more difficult. To equalize the difficulty, we apply a dimension reduction  (Probabilistic Principal Component Analysis~\cite{tipping1999probabilistic}) to $d=128$ (FEI), $256$ (CFP), and $512$ (LFW).
The parameters of PPCA are learned on a different set of images, not on the enrolled templates and queries.
The vectors are also $L_{2}$ normalized.
With such post-processing, the average correlation between positive pairs equals 0.83 (FEI), 0.78 (CFP), and 0.68 (LFW) with a standard deviation of $0.01$. Despite the dimension reduction, the hardest dataset is LFW and the easiest FEI.


In one simulation run, the enrollment phase makes random groups with the same number $n$ of members.
A user claims she/he belongs to group $g$.  This claim is true under hypothesis $\Hyp_{1}$ and false under hypothesis $\Hyp_{0}$ (\ie\ the user is an impostor).  Her/his template is quantized to the sequence $\Q$, and $(\Q, g)$ is sent to the system, which compares $\Q$ to the group representation $\Y_{g}$. This is done for all impostors and all queries of enrolled people. One Monte-Carlo simulation is composed of $20$ runs.
The figure of merit is $\Pfn$ when $\Pfp = 0.05$.

\subsection{Exp. \#1: Comparison to the baselines}
\label{fig:exp1}
Our scheme is compared to the following baselines:
\begin{itemize}
	\item EoA-SP and AoE-SP~\cite{Gheisari2019icassp} (signal processing approach)
	\item EoA-ML and AoE-ML~\cite{Gheisari_2019_CVPR_Workshops} (machine learning approach)
\end{itemize}
The drawback of these baselines is that the length $m$ of the data structure is bounded. Here, it is set to maximum value, \ie\ $m=d$ the dimension of templates.

Our scheme allows more freedom. 
Setting $m=8\times d$ produces a much bigger representation.
It is not surprising that our scheme is better than the baselines.
Fig.~\ref{fig:baseline} validates our motivation to get rid off the drawback of the baselines with limited $m$, to achieve better verification performance.
These results are obtained with the dense solution.
Indeed, despite all our efforts, we could not achieve better results with the sparse solution.
This confirms the lesson learnt from Fig.~\ref{fig:MIcor}: the dense solution outperforms the sparse solution when the average correlation between positive pairs is lower than $0.95$.

The improvement is also better as the size of groups increases.
We explain this by the use of the types, \ie\ $Y = T$.
Equation~\eqref{eq:Cgauss} shows that $\fC$ increases with $n$ for the dense solution, compensating for aggregating more templates.


 

\subsection{Exp. \#2: Reducing the size of the group representation}
There are two ways for reducing the size of the group representation.
The first means is to decrease $m$, the second means is to lower $\fC$ thanks to a surjection.
Sect.~\ref{sub:surjection} presented optimal surjections from $\mathcal{T}_{2,n}$ to $\Alpy = \{0,1\}$.
We found experimentally good surjections to sets $\Alpy$ for $|\Alpy| \in \{3,4,8\}$.

\begin{figure}[b]
	
	\centering	
	\includegraphics[width=0.85\columnwidth]{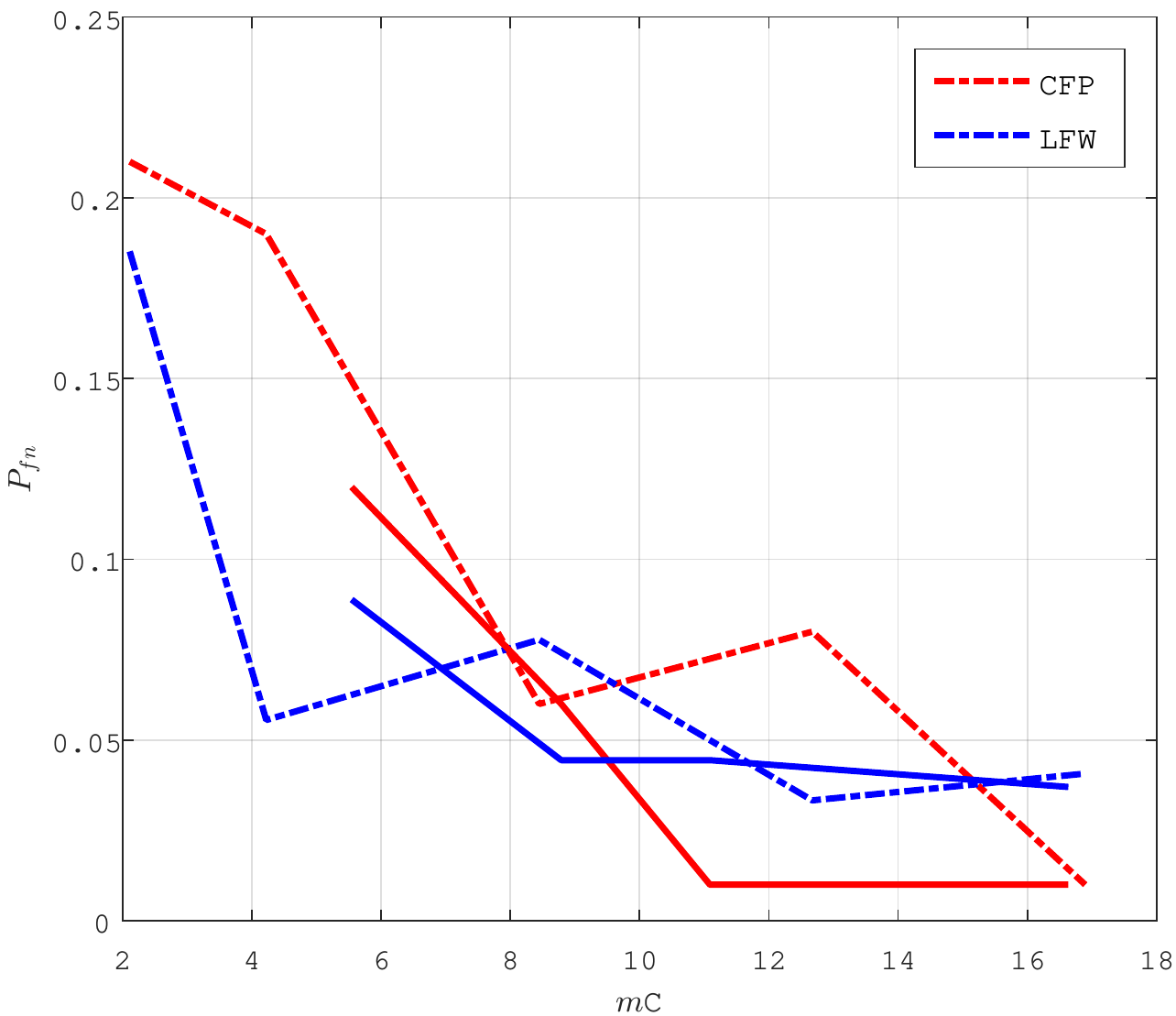}
	\caption{Verification performance $\Pfn@\Pfp=0.05$ \textit{vs.} $m\fC$, for $n=16$. This quantity is reduced by decreasing $m$ (dashed lines) or by decreasing $\fC$ thanks to a surjection (solid lines).}
	\label{fig:elle}%
\end{figure}

This is done according to the following heuristic. Starting from $\mathcal{T}_{2,n}$, we iteratively decrease the size of $\Alpy$ by one. This amounts to merge two symbols of $\Alpy$. By brute force, we analyse all the pairs of symbols measuring the loss in $\fV$ induced by their merging. By merging the best pair, we decrease the number of symbols in $\Alpy$ by one. This process is iterated until the targeted size of $\Alpy$ is achieved. This heuristic is not optimal, but it is tractable. 
Fig.~\ref{fig:elle} compares these two means. Employing a coarser surjection is slightly better in terms of verification performances. 
 
\subsection{Unexpected results}
We have argued that FEI $<$ CFP $<$ LFW in terms of difficulty due to the opposite ordering of the datasets typical correlation $c$ between positive pairs. Eq.~\eqref{eq:eta0} shows that a lower $c$ produces a higher $\eta_0$ (and $\eta_1$), whence a lower $\fV$.
In Fig.~\ref{fig:baseline}, the experimental results contradict this intuition. 

This may be explained by the Signal to Noise Ratio at the template level. We define it as $c^2 / v_0$ where $c$ is the average correlation for positive pairs and $v_0$ is the variance of this correlation for negative pairs. If a negative query is uniformly distributed over the hypersphere, then its correlation with an enrolled template is approximatively distributed as a centered Gaussian distribution with variance $v_0 = 1/d$.

Yet, $d$ has no impact on $p$, $\eta_0$, and $\eta_1$.
We suppose that its impact is tangible on the entropy of the template vectors.
Sect.~\ref{sec:Setup} assumes that the enrolled sequences are statistically independent.
This assumption is not granted with the embedding of Sect.~\ref{sec:Real}.
Yet, a bigger $d$ favors the independence (or at least the decorrelation) between real template vectors.




\section{Conclusion}
Our theoretical study justifies that the dense setup is more interesting in terms of verification performance $\fV$ and security level $\fS$ unless we are operating in the high-SNR regime where the positive queries are very well correlated with the enrolled templates. This statement holds for any embedding, yet some are certainly more suited than others depending on $d$, $c$, and the geometrical relationship among positive pairs.


\section*{Acknowledgment}
This work is supported by the project CHIST-ERA ID\_IOT 20CH21 167534.
\appendix
Let us first explain how $\fV$ is computed.
Denote $P_i(q,y):= \Pr(Q=q,Y=y|\Hyp_i)$ and
channel $W(q|x):= \Pr(Q=q|X=x)$, $\forall y\in\Alpy, q\in\Alp$ and $i\in\{0,1\}$.
Then,
\begin{equation}
\fV = \sum_{q,y} P_1(q,y)\log\frac{P_1(q,y)}{P_0(q,y)},
\end{equation}
with $P_0(q,y) = \Pr(Q=q)\Pr(Y=y)$ and
\begin{equation}
P_1(q,y) = \sum_{x\in\Alp} \Pr(Y=y,X=x) W(q|x).
\end{equation}

\subsection{Surjection to $\Alpy=\{0,1\}$}
\label{app2}
We assume here the noiseless setup allowing to write $\Pr(Y=y,X=x)$  as $P_1(x,y)$.
Inspired by traitor tracing, we consider a probabilistic surjection where $\Pr(\fun{r}(t)=1) = \theta_t$.
The vector $\thetab\in[0,1]^{n+1}$ parametrizes the surjection.
Denote by $\nabla_{\thetab} \fV(t)$ the derivative w.r.t. $\theta_t$.
After some lengthy calculus:
\begin{eqnarray}
\nabla_{\thetab} \fV(t)&=&n^{-1}K_{1}(p,\thetab)(t - nK_{2}(p,\thetab)),\\
\label{eq:DerivMutInfoKnowing2}
K_{1}(p,\thetab) &=& \Pr(T=t)\Delta,\nonumber\\
K_{2}(p,\thetab) &=& \frac{\fun{h}^\prime(P_1(0,1)) - \fun{h}^\prime(\Pr(Y=1))}{\Delta}, \nonumber\\
\Delta &=&\fun{h}^\prime(P_1(0,1)) - \fun{h}^\prime(P_1(1,1)).\nonumber
\end{eqnarray}
It is not possible to cancel the gradient $\nabla_{\thetab} \fV$.
The optimal $\thetab$ thus lies on the boundary of the hypercube $[0,1]^{n+1}$.
This makes the surjection deterministic.
Assuming $\Pr(Y=1|X=0) < \Pr(Y=1|X=1)$, then $0<K_{1}(p,\thetab)$ and $0<K_{2}(p,\thetab)\leq1$ because $\fun{h}^\prime(\cdot)$ is strictly decreasing.
This makes $\nabla_{\thetab} \fV(0)<0$ and $\theta_{0}$ must be set to the lowest possible value, \ie\ $\theta_{0}=0$, to increase $\fV$ at most. This is indeed the case for any $\theta_{t}$ with $t<K_{2}(p,\thetab)$.
In the same way, $\theta_{n}=1$ and so is $\theta_{t}$ if $t>K_{2}(p,\thetab)$.
Yet, for a given $\thetab$, $K_{2}(p,\thetab)$ ranges from 0 to $1$ as $p$ increases from 0 to 1.
Therefore, $\thetab=(0,\ldots,0,1,\ldots,1)$ is optimal only over an interval of $p$.

For $n$ odd and $p=1/2$, $\theta_t = 0$ if $t\leq(n+1)/2$, and 1 (\ie\ majority vote) otherwise is optimal because $K_2(1/2,\thetab) = 1/2$ ($\Pr(Y=1)=1/2$ and $P_1(0,1) = 1 - P_1(1,1)$).

The `All-1' surjection: $\thetab =(0,1,\ldots,1)$ makes $P_1(1,1) = 1$ so that $\nabla_{\thetab} \fV(t) = +\infty$ if $t>0$ and $<0$ for $t=0$.

\subsection{Impact of the channel}
\label{app3}
Suppose that $\eta$ is a parameter of the channel $W(\cdot|\cdot)$.
Then
\begin{equation}
\label{eq:dVdeta}
\frac{\partial\fV}{\partial\eta} = 
\sum_{q,y} \frac{\partial P_1(q,y)}{\partial\eta}\log\frac{P_1(q,y)}{P_0(q,y)},
\end{equation}
because $\sum_{q,y}\frac{\partial P_1(q,y)}{\partial\eta} = \frac{\partial\sum_{q,y} P_1(q,y)}{\partial\eta}=0$ and
$\sum_{q,y}\frac{P_1(q,y)}{P_0(q,y)}\frac{\partial P_0(q,y)}{\partial\eta} =
\sum_{q}\frac{\partial \Pr(Q=q)}{\partial\eta} = 0$.

Suppose now that $\eta = \eta_0 := W(q|0),\forall q\in\Alp\backslash{0}$.
Then,
\begin{equation}
\frac{\partial P_1(q,y)}{\partial \eta_0} = \Pr(X = 0, Y = y) \,\forall q\in\Alp\backslash\{0\}.
\end{equation}
Taking~\eqref{eq:dVdeta} around the noiseless channel where $\eta_0 = 0$ and $\Pr(X = 0, Y = y) = P_1(0,y)$ because $Q=X$:
\begin{equation}
\left.\frac{\partial\fV}{\partial\eta_0}\right|_{\eta_0=0}=
\sum_{y,x\neq 0}P_1(0,y) \log\frac{P_1(x,y)}{P_0(x,y)} + \ldots
\end{equation}
We only express the first terms to outline that if $P_1(x,y) = 0$ while $P_1(0,y)$ and hence $P_0(x,y)$ are not null, then this derivative goes to $-\infty$. A small deviation from the noiseless case with $\eta_0\neq 0$ has a major detrimental impact on $\fV$.
That situation happens for sure when working with type, \ie\ $Y = T$:
Consider the null type $t_0$ obtained when $X_1 = \ldots = X_n = 0$: $P_1(0,t_0)>0$ while $P_1(x,t_0) = 0$, $\forall x\neq0$.

One can prove that the surjection can mitigate this effect if $\exists t\neq t_0: \fun{r}(t) = \fun{r}(t_0)$ and $P_1(0,t)>0$.
This happens with the majority vote of the dense setup, but unfortunately, not with of the `All-1' surjection in the sparse setup.



\bibliographystyle{IEEEtran}
\bibliography{ms}

%
%
%
%
%

\end{document}